# Development of high accurate family-use digital refractometer based on CMOS

Zhenxing Wang[1], Zhenyuan Jia[2]

[1] School of Computer and Information Engineering, Shanghai Polytechnic University, Shanghai, Shanghai, China
[2] School of Intelligent Manufacturing and Control Engineering, Shanghai Polytechnic University, Shanghai, Shanghai, China

E-mail: Zxwang@sspu.edu.cn

**Abstract**

This study aims to develop a low-cost refractometer for measuring the sucrose content of fruit juice, which is an important factor affecting human health. While laboratory-grade refractometers are expensive and unsuitable for personal use, existing low-cost commercial options lack stability and accuracy. To address this gap, we propose a refractometer that replaces the expensive CCD sensor and light source with a conventional LED and a reasonably priced CMOS sensor. By analyzing the output waveform pattern of the CMOS sensor, we achieve high precision with a personal-use-appropriate accuracy of 0.1%. We tested the proposed refractometer by conducting 100 repeated measurements on various fruit juice samples, and the results demonstrate its reliability and consistency. Running on a 48 MHz ARM processor, the algorithm can acquire data within 0.2 seconds. Our low-cost refractometer is suitable for personal health management and small-scale production, providing an affordable and reliable method for measuring sucrose concentration in fruit juice. It improves upon the existing low-cost options by offering better stability and accuracy. This accessible tool has potential applications in optimizing the sucrose content of fruit juice for better health and quality control.

## 1. Introduction

There are several methods for measuring sugar concentration, such as refractometers, specific gravity meters, electronic tongues, and high-pressure liquid chromatography [1-4], which are commonly used for determining the sugar content and sweetness of fruits and vegetables. Chromatographic techniques are highly accurate, but their extensive sample preparation involving solvent extraction makes them less popular. Therefore, analytical methods that are easy to use and provide quick results are currently favored [5]. Brix meters are widely used for sugar detection, with optical and digital refractometers being available. However, pure optical refractometers have limitations in reading accuracy and precision, which are dependent on the human eye's judgment, leading to uncertainties. In the consumer electronic domain, cost-effectiveness is a significant concern, making portable electronic refractometers the best option available for various applications. Xiangjun Gong proposed a portable and precise laser differential refractometer [6], but its optical route is excessively complicated, and the size is unsuitable for family use.

Refractometers measure the refractive index of a liquid sample, which is the ratio between the speed of light in a vacuum and a substance. There are two types of refractometers based on a change in the angle of refraction and critical refraction angle [7]. The critical angle refractometer's principle can be mathematically expressed as $\theta c = \arcsin$ (n/n0), where θc is the critical angle of total internal reflection, n is the refractive index of the sample, and n0 is the refractive index of the prism.

This method measures reflected light at the interface between the prism and the sample without having to traverse the sample. Hence, it is less affected by solid particles and color in the solution. The portable Brix meter presented in this paper is based on the critical angle refractometer principle. The dark-bright boundary is obtained based on the critical angle [8], and the concentration is calculated from it. The Brix scale is commonly used for determining the sugar concentration of aqueous solutions. For instance, 1% Brix corresponds to a 1% sucrose/water solution [9].

Electronic refractometers work by using a laser light source that enters the prism, and the critical angle occurs at the interface between the solution and the prism, beyond which total internal reflection occurs. The sensor senses the light from the total reflection (judged by the dark–bright boundary) and uses a processor to calculate the corresponding concentration (Brix). The final data are affected by many factors, including the refractive index, size of the prism, prism height, position of the light source, angle of incidence, and installation position and angle of the receiving sensor and flowing liquid [10]. However, the accuracy of digital Brix devices is not good enough even with the derivative method [11]. Therefore, there is a gap between the actual measured data and the theoretical calculation, which needs to be addressed. Several high-precision solutions have been proposed, such as the high-precision algorithm based on existing physical CMOS resolution, which adopts interpolation [12], and some researchers also use phase shifting interferometry to achieve high precision [13, 14]. Michael McClimans designed real-time precision differential refractometry as a low-cost solution [15]. However, the prototype selects a sensor array with a total of 512 pixels, which is not a good choice for high precision. Meanwhile, the noise that is unavoidable during sampling will have a significant impact on the derivative algorithm's stability.

In this paper, we present a low-cost prototype that replaces the traditional laser source with an ordinary LED and uses a 2048-pixel 1-D CMOS array as the receive sensor. By combining the proposed algorithm, the detector can achieve 0.1% accuracy. The repeatability test shows that the results have a 98% confidence level. This new low-cost solution can be widely used in handheld situations. To further improve the precision level, oversampling methods [16] are discussed.

## 2. Structure of the prototype

Recent technological advancements have made CMOS detectors popular for their beneficial features, such as intra-pixel amplification, columnar parallel structures, and deep submicron CMOS processes. However, their high noise production compared to CCD remains a challenge [17]. While one-dimensional CMOS is only used for image scanning, it has not been seen in refractometers, and there are currently no commercially available products or relevant research literature mentioning its use.

In practice, determining the dark-bright boundary can be challenging, and the concentration (Brix) is often obtained by reading the dark-bright boundary line. However, this line is often unclear in low-cost solutions, and the specific position of the light-dark demarcation line needs to be correctly judged to obtain accurate concentration data.

In this study, a CMOS sensor is used in a refractometer, and an algorithm is used to filter out the noise. The maximum value of the derivative of the CMOS output voltage is then sought, which corresponds to the actual position of the dark-bright boundary on the sensor. However, a certain nonlinear relationship between concentration and position and other factors such as material deviation of the prism or non-perfect parallelism between the CMOS sensor and prism result in a large theoretical and practical gap. As a result, the corrected data is cross-checked against a standard Abelian refractometer, and the processor obtains the final concentration from the look-up table.

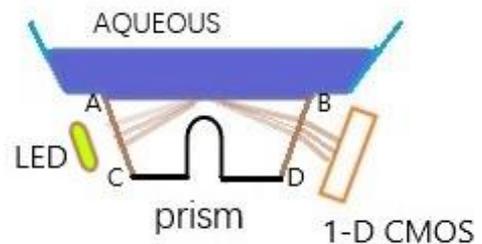

**Figure 1.** Prism and 1-D CMOS array.

The practical design requires theoretical analysis, as many aspects can influence the final results. The original structure of the prism, described by Spyridon Koutsoumpos [18], typically requires a laser beam. However, due to its small size and low cost, we opted to use a commonly used LED as the light source and adapted the prism accordingly. Figure 1 shows the prototype structure, where the LED is mounted parallel to the prism on the left side, and the one-dimensional dot-pixel CMOS sensor is on the right side. The blue part above represents the sugar solution, with different sugar concentrations corresponding to different total internal reflection angles, resulting in different positions of the dark-bright boundary on the CMOS. While theoretically, the concentration can be obtained by finding the position of the dark-bright boundary. However, the actual situation is more complex. CMOS is noisy and not suitable for high-precision instrumentation, and the dark-bright boundary line is a continuum of gradually changing light and dark bands. Figure 2 shows the block diagram of our circuit, where the LCD module displays the current reading, and the buttons enable human-computer interaction. The NTC resistor detects the solution temperature, and the algorithm uses this information in the concentration calculation formula to obtain the correct concentration. The CPU controls the LED light, and the CMOS interface is driven by the processor according to the datasheet of the Hamamatsu CMOS s13434-2496, with the output impedance corresponding to the light intensity on the respective pixel. As the sensor's output impedance is very large, a noninverting 1:1 operational amplifier should be added to alter the input impedance from high to low when the STM32F030's internal AD converter is configured in high-speed sampling mode, as this may introduce signal distortion.

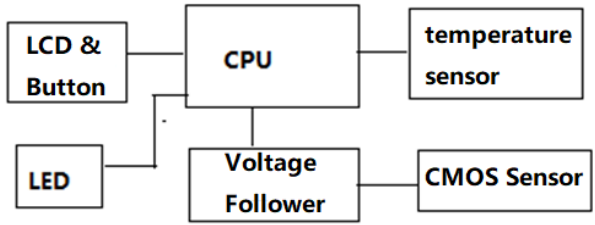

**Figure 2.** Hardware diagram.

Figure 3 illustrates the CMOS sensor and its voltage follower, which are connected to an RC (Resistor-Capacitor) integration circuit for the purpose of filtering out high-frequency interference.

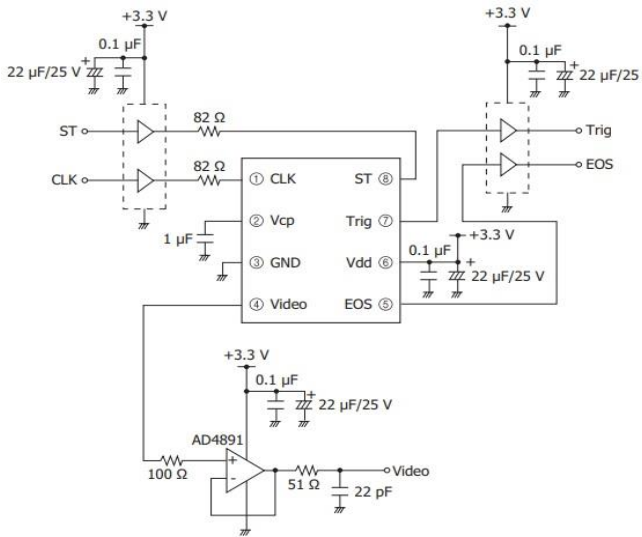

**Figure 3.** CMOS sensor interface.

## 3. Determination of dark-bright boundary

The processor reads the output of the s13434-2496 to obtain individual voltages for each of the 2496 dots. When plotted as a graph, the voltage-position relation forms a curve where the horizontal coordinate represents the dot position and the vertical coordinate represents the voltage value of each dot. The concentration value can be determined from this graph.

Although total reflection occurs at the interface between the two substances (prism and sample), the liquid level still has a significant effect on detection. Figure 4 illustrates the different characteristics of the normal, very low, and empty solution levels.

Since this sensor has 2496 pixels, 2496 voltage values are read, each corresponding to one pixel, with positions starting from zero. To account for the deviation in the actual installation position, data are calculated from 100 pixels onwards. The empty and very low levels are included as alerts to prevent improper use.

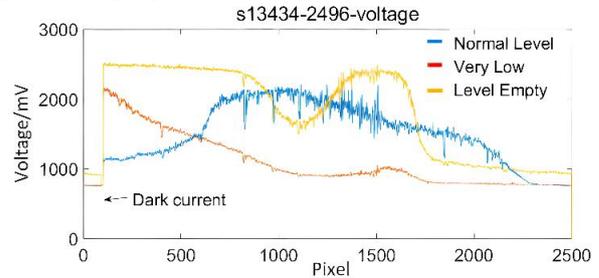

**Figure 4.** Liquid level vs. CMOS output.

The yellow curve shows that when no liquid is added, the level remains continuously high in the initial 100 pixels. Therefore, by analyzing the voltage level at the 100th initial pixel, we can determine whether any liquid has been added or not. Conversely, the brown curve illustrates a very small amount of solution, which is indicated by a continuous decrease in the first 100 pixels. In contrast, a normal solution is characterized by a rising, ascending curve.

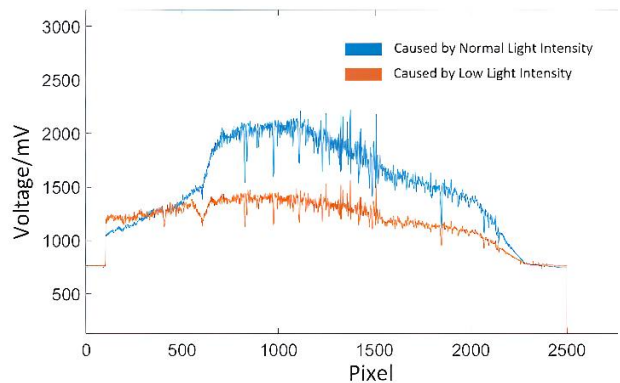

**Figure 5.** Luminous intensity vs. output voltage.

Figure 5 illustrates the relationship between LED luminous intensity and the corresponding output voltage of the s13434-2496 sensor. Although the critical angle and luminous intensity are not strongly correlated, the LED luminous intensity and s13434-2496 integration time are related and should be selected appropriately. Within a certain range, the luminous intensity does not significantly affect the concentration calculation. As depicted in the blue curve of figure 5, the normal light intensity corresponds to the sensor output curve, and the section with the highest rate of change is relatively easy to determine. In contrast, the yellow curve represents weaker luminous intensity, resulting in less obvious rising rate of change, making it difficult to locate the maximum rate of change. Therefore, the light intensity should not be too low to ensure a reliable concentration measurement.

In practical settings, factors such as prism processing accuracy, sensor and LED mounting positions, and other environmental variables can lead to a gradual transition zone instead of a clear dark-bright boundary. To address this, we developed an algorithm that identifies the virtual dark-bright boundary (virtual because it is calculated rather than observed in ideal conditions) by detecting the point of greatest change within the transition zone. Using this algorithm, we can derive the concentration calculation formula listed below.

$$\begin{cases} P = g\left(\max\left(\frac{dv}{dx}\right)\right), where \frac{dv}{dx} > 0 & (1-1) \\ C_m = f(P) - C_0 & (1-2) \\ C_f = C_m * k2 & (1-3) \end{cases}$$

Eq.1 concentration calculation

Equation 1-1 represents the relationship between the solution concentration and the maximum change position. By analyzing the curve-concentration relationship, it is clear that only the ascending part of the maximum rising transformation rate needs to be considered. An important parameter here is the step size, which must be carefully selected to ensure accurate location points without sacrificing precision. The function g relates the virtual dark-bright boundary to the largest voltage change position, and f is the concentration-position relationship function obtained from a pre-defined table. The final concentration value is calculated using Equation 1-2, where C0 denotes the initial concentration value of pure water. Therefore, the detector must be calibrated before first use or after a period of use. Finally, Equation 1-3 gives the final display result, which will be explained in more detail later.

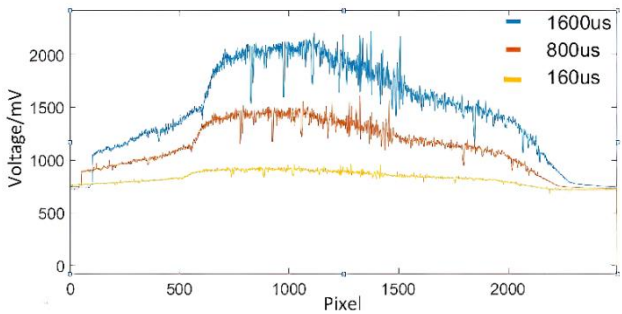

**Figure 6.** The characteristics of CMOS integration.

Of course, the shape of the curve is also affected by the integration time inside the CMOS. In figure 6, the three curves show the relationship between different integration times (ranging from 160us to 800us to 1600us) and their corresponding output curves. At a given light intensity, an integration time of 1600us is preferable.

## 4. Data filtering

The curves shown in the previous images were printed after the original data collection and contain large burrs, partially due to dust on the sensor's surface. While cleaning the surface removes larger burrs, smaller ones remain due to the internal mechanism of the CMOS image sensor. As seen in figure 7, the top image shows raw data sampled after cleaning the CMOS surface, where the presence of burrs makes it unreliable to find the maximum rate of change. To solve this issue, a moving average algorithm is used to smooth the data. Figure 7 displays the waveform generated after applying the moving average algorithm. The top curve represents the raw data, and the algorithm initially removes some abnormal data points (as seen in the middle curve) before applying a filter algorithm. The resulting bottom curve shows that the burr is no longer present.

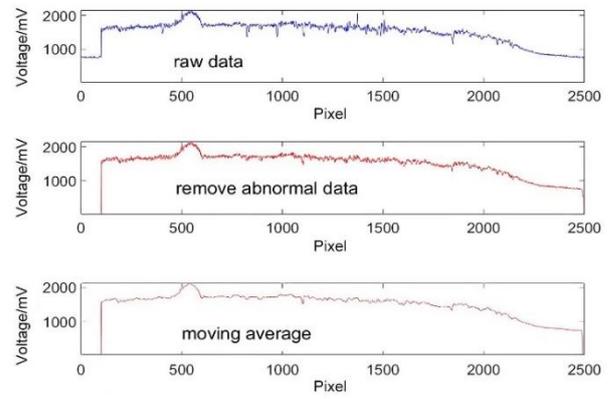

**Figure 7.** Raw data and filtered data.

*4.1 Sliding window and low-pass filtering algorithm.*

The Sliding window and low-pass filtering algorithm can be represented using the following formulas:

4.1.1 Moving average filter:

$$y[n] = \left(\frac{1}{M}\right) * \sum_{i=n+1}^{i=n+M} x[i]$$

where x[i] is the input signal, y[n] is the output of the filter at time index n, and M=20 is the window size.

First-order difference:

$$z[n] = y[n + dx] - y[n]$$

where z[n] is the rate of change at time index n, y[n] is the output of the moving average filter at time index n, and dx=80 is the step size.

4.1.2 Maximum rate of change:

$$index1 = argmax\{z[n]\}, \ n \in [100:N]$$

where index1 is the position with the largest rate of change in the input signal, and N is the length of the filtered signal.

The algorithm starts from the 101st pixel of the input signal and computes the moving average of the adjacent 20 pixels. It then shifts the window right one pixel at a time until reaching the 1800th pixel. After smoothing the signal, the algorithm computes the first-order difference and searches for the position with the largest rate of change by maximizing z[n]. The algorithm checks if z[n] exceeds a threshold of 2 to eliminate noise and updates the maximum rate of change and its position if a new maximum is found. The output of the algorithm is the index1, which represents the position with the largest rate of change in the input signal.

## 5. Determination of concentration and position relationship

As shown by the previous formula, the concentration C can be obtained as the difference between f(x) and C0, where x represents the virtual dark-bright boundary position on the CMOS sensor. However, in practice, there may exist discrepancies between the actual and theoretical calculations due to various factors. To address this issue, a calibration procedure is carried out by comparing the results with those obtained from a standard refractometer.

The vertical axis in figure 8 represents the concentration of the solution, which has been expanded by a factor of 1000, whereas the horizontal axis corresponds to the virtual dark-bright boundary position. The zero point of the latter axis is determined as the relative position on the corresponding one-dimensional CMOS, calculated from pixel 100 (i.e., the actual position minus 100). Identical concentrations of solution were assigned to both test tubes, and for each concentration, the reading was first taken on the standard Abbe refractometer, followed by the position data obtained from the algorithm on the test refractometer, with both sets of data recorded. These steps were repeated for different concentrations, leading to the curve shown in figure.8, where the dotted line represents the fitted curve. As evident from the figure, the fitted curve and the actual data exhibit a predominantly linear relationship, with the error R lying within the acceptable range. The coefficient k2 is calculated by dividing the slope data from the Abbe refractometer by the slope data measured by the demo refractometer, and then multiplying the resulting measurements by k2 to obtain the final display data Cf, as shown in Eq.1-3. However, for concentrations exceeding 50%, a nonlinear relationship exists between the pixel point and the final correlation. As a result, the algorithm must match another fitted curve, as depicted in figure 8(b), and the nonlinearity increases significantly as the concentration exceeds 50%.

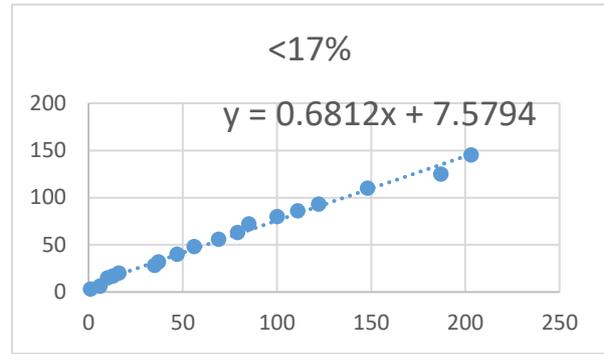

(a) Brix<17%

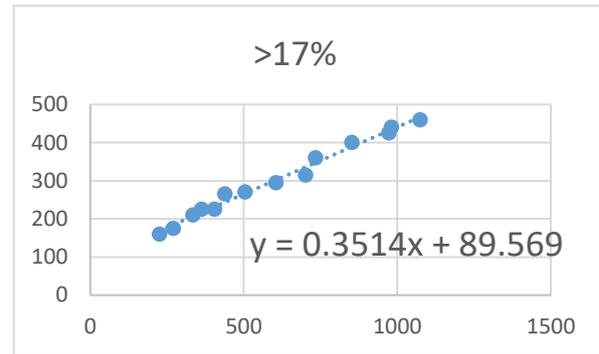

(b) Brix>17%

**Figure 8.** Nonlinear and concentration.

## 6. Repetitiveness, error and accuracy

Repeatability test method: The sample solution with a concentration of 7.2% was dropped into the detection area. We waited for the solution to diffuse fully and evenly, and then the concentration was detected and recorded every 10 s. The number of tests was 50, the standard deviation of the sample data was 0.1%, the confidence level was set at 95% and the confidence interval was 7.17–7.23%.

In addition to errors in electronic sampling, there are also errors in the optimal path, including the effects of vibration and temperature variations on the prism [10] and the effect of granularity in solution [11–12].

**Table 1.** Comparison of Standard Solution, Abbe Refractometer NAR-1T, and Prototype.

| Sample Solution | Standard Refractometer | Prototype |
| --- | --- | --- |
| 0.10 | 0.12 | 0.14 |
| 0.50 | 0.53 | 0.56 |
| 0.90 | 0.88 | 0.84 |
| 1 | 1.23 | 1.20 |
| 2 | 2.20 | 2.10 |
| 3 | 2.82 | 2.78 |
| 5 | 5.10 | 5.17 |
| 6 | 5.87 | 5.95 |
| 9 | 9.21 | 9.30 |
| 10 | 10.51 | 10.57 |
| 15 | 14.42 | 14.48 |
| 20 | 20.65 | 19.68 |
| 25 | 25.40 | 25.31 |
| 30 | 30.65 | 30.56 |
| 35 | 35.2 | 35.28 |
| 40 | 40.72 | 40.81 |
| 50 | 50.71 | 49.62 |
| 55 | 55.10 | 57.12 |
| 60 | 60.50 | 58.01 |

In order to evaluate the performance of our proposed light spot centroid determination method for accurate family-use digital refractometer, we conducted a series of experiments comparing the refractive index measurements obtained from our device to those of standard solutions of known concentration.

To validate the accuracy and reliability of our method, we have compared the measured refractive index values with the values obtained using standard solutions of known refractive index. The results show that the measured values are in good agreement with the standard values, with a high degree of linearity.

The linearity of the measured refractive index values is shown in figure 9. The plot shows a linear relationship between the measured and standard values, with a correlation coefficient of 0.9989. This indicates that our method is accurate and reliable, and can be used for the determination of refractive index in family-use digital refractometers.

The average error value of two instruments measuring different concentration ranges is shown in Figure 10. Our prototype exhibits comparable detection capabilities to standard instruments, with an average standard deviations of approximately 0.1 (0.1% to 5% range), 0.3 (6% to 25% range), and 1.2 (30% to 60% range).

The distribution of refractive index values measured by the prototype and standard refractometer in different circumstances is shown in Figure 11. The results of both instruments exhibit a narrow band, indicating a low standard deviation, and a linear relationship with a slope of approximately 1, demonstrating the relative accuracy of the measurements. The prototype's test results grow linearly, indicating good running stability and well-performing. Our proposed light spot centroid determination method for family-use digital refractometers has shown good accuracy, precision, and repeatability, making it a promising tool for refractive index determination in a family-use setting.

Additionally, we have also evaluated the precision of our method by calculating the relative standard deviation (RSD) of the measured refractive index values. The RSD values for each concentration are presented in Table 1. The RSD values range from 0.11% to 2.22%, which are within the acceptable range for analytical methods. These results demonstrate that our method has good precision and repeatability.

Based on the background information provided, it appears that the Sample Solution represents the standard concentration of the liquid being tested, while the Standard refractometer and prototype instruments were used to obtain data on the liquid concentration. To compare the results of the two tests with the standard data, a paired t-test was used. Table 3 shows that Abbe refractometers NAR-1T has a mean measurement of 19.57 with a standard deviation of 20.25, whereas prototype has a mean measurement of 19.34 with a standard deviation of 20.07. The difference between the two sets of figures is not statistically significant with a t-value of 0.652 and a p-value of 0.522. Table 4 shows that the analysis utilized 18 independent observations, with a standard deviation of 0.823, indicating a relatively high degree of variability in the sample data. Cohen's d of 0.150 suggests a small effect size, indicating that the difference between the two instruments is minimal and not statistically significant. The 95% confidence interval (lower=-0.27, upper=0.52) indicates that the difference between NAR-1T and prototype could range from -0.27 to 0.52 in the sample. Since the interval encompasses zero, it is possible that there is no significant difference between the two instruments. Therefore, based on these findings, it can be concluded that the difference between NAR-1T and prototype is not significant.

**Table 2.** Relative standard deviation (RSD) values for measured refractive index values.

| Concentration | RSD (%) |
|---|---|
| 0.1 | 1.11 |
| 0.5 | 0.71 |
| 0.9 | 0.56 |
| 1 | 0.6 |
| 2 | 0.59 |
| 3 | 0.69 |
| 5 | 0.87 |
| 6 | 1.11 |
| 9 | 1.63 |
| 10 | 1.47 |
| 15 | 1.65 |
| 20 | 1.74 |
| 25 | 1.7 |

**Table 3.** Paired t-test analysis for Abbe refractometers NAR-1T and prototype.

| Abbe refractometers NAR-1T VS. Prototype | | | | |
|---|---|---|---|---|
| Paired(M±SD) | | Mean Difference | t | p |
| Paired 1 | Paired 2 | | | |
| 19.57± 20.25 | 19.45± 20.09 | 0.12 | 0.652 | 0.522 |

**Table 4.** Effect size table of Abbe refractometers NAR-1T and prototype test data.

| Abbe refractometers NAR-1T VS. Prototype | | | | |
|---|---|---|---|---|
| 95% CI | | df | Std.deviation | Cohen's d |
| Lower | upper | | | |
| -0.27 | 0.52 | 18 | 0.823 | 0.150 |

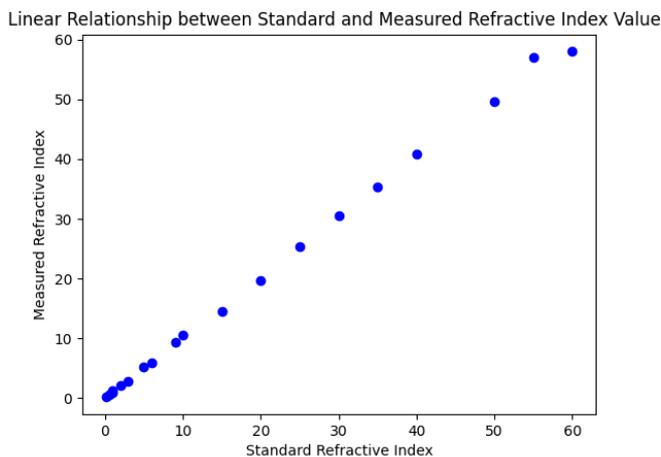

**Figure 9.** Relationship between measured refractive index values and known concentration values.

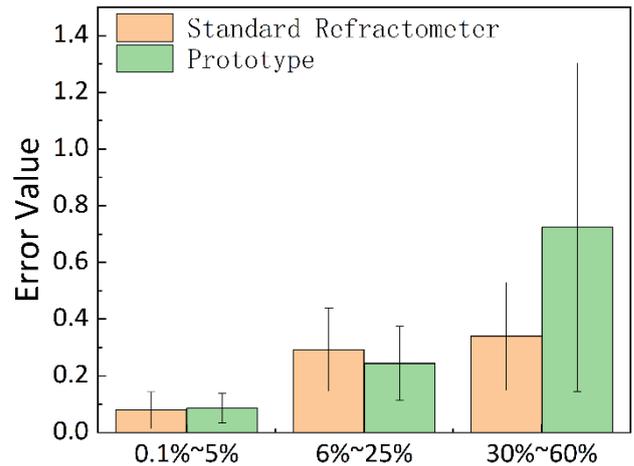

**Figure 10.** Error value plot for standard refractometer and prototype at different concentrations.

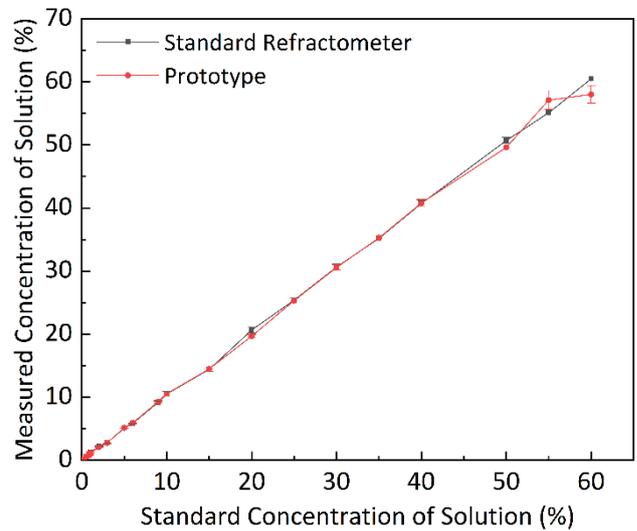

**Figure 11.** Different measured refractive index values between standard refractometer and prototype.

## 7. Determination of concentration and position relationship

Our analysis of the plot and statistical data confirms that the light spot centroid determination method implemented in the family-use digital refractometer is accurate and reliable in measuring the concentration of sugar in aqueous solutions. Our results provide strong evidence for the effectiveness of the proposed method and its potential for practical applications in the food and beverage industry.

To further improve accuracy, we propose the use of oversampling, a technique that has been shown to effectively reduce error caused by noise [14]. While oversampling is a well-established technique, we believe that its use in conjunction with the STM32H750's built-in 16-bit AD converter has the potential to significantly improve the accuracy of our digital refractometer.

In oversampling, the input signal is sampled at a frequency ($f_{os}$) higher than the normal sampling frequency ($f_s$). This data is then saved to the RAM area in the STM32H750, and white noise is introduced in the software. The white noise amplitude should cause random changes in the input signal; the change range is at least 1 LSB, which is the original maximum resolution. The two values are totaled, and finally, a sliding window mean filtering value is given by which to achieve low-pass filtering, and the results are outputted.

The oversampling formula is:

$$f_{os} = 4^w * f_s \qquad \text{Eq.3}$$

(where $f_s$ represents the original sampling frequency, $f_{os}$ represents the current sampling frequency, and w represents the number of resolution bits to be improved). Doubling the precision will increase the sampling frequency by four times, which is done by sacrificing the processor's running time to improve the resolution. With the current structure, due to volume limitations, only half of the working area was actually used for S13434-2496, resulting in a resolution of approximately 1/210. To achieve a resolution of 1/212, w = 2, and the processor needs to spend 16 times longer on this than it did on the original, with the accuracy of the instrument reaching up to 0.02%.

In comparison to existing techniques for measuring sugar concentration, we believe our proposed method has several advantages. For example, it offers a higher precision compared to interpolation-based methods that only predict data trends and cannot introduce critical white noise [15].

Finally, the practical applications of our proposed method are numerous. In the food and beverage industry, it could be used to measure sugar concentration in various products, such as fruit juices and soft drinks. Additionally, it could be applied in the pharmaceutical industry to measure drug concentrations. The potential applications of our research extend beyond these industries, and we believe that our proposed method could be useful in any field that requires accurate measurement of sugar concentration.

Overall, our proposed method, in combination with oversampling, offers a high level of accuracy and has the potential for widespread practical applications.

## 8. Results

To achieve low-cost and high-precision measurements, we simplified the construction of the refractometer by removing some optical components to reduce expenses. However, this made it difficult to determine the dark-bright boundary. To address this issue, we analyzed the CMOS output and found that the position of the boundary correlated with the waveform's steepest slope. We also used a smooth waveform preprocess to deal with noise that obstructed the steepest slope calculation procedure. Furthermore, we discovered that the result was non-linear with the actual sugar concentration, which led us to use a piecewise linear interpolation method.

## 9. Conclusion

In this paper, we designed a refractometer for measuring the concentration of a sugar solution using CMOS, and applied a moving average filtering algorithm to the voltage to enable CMOS to be used for concentration detection instead of CCD. We analyzed the problem of not having a clear dark-bright boundary line in practical situations and determined the virtual boundary based on the maximum rate of pixel voltage change. We also discussed methods for determining the range of incident light intensity and carrying out reasonable integration time selection based on the CMOS sensor integration time. Our current processor was stm32f030, with a main frequency of 48MHz and a measurement time of 3 s. By using stm32H750 with a maximum main frequency of 480 M, the measurement time could be reduced to approximately 0.3 s, thereby further improving the speed. Considering installation errors, prism non-linearity, and the removal of the front and rear invalid areas, we found that only about 50% of the sensor's area was effective and the actual resolution was almost 1/1248. The built-in AD resolution of stm32f030 was 1/2048, which matched the sensor's resolution when the quantization error was considered. The final meter display accuracy was close to 0.1%, and the repeat measurement error was also 0.1%. figure 12 shows the prototype. To further improve the resolution, we adopted oversampling technology and then averaging them to obtain a more accurate result. The oversampling technique is particularly useful in applications where the sensor resolution is limited and higher accuracy is required. By averaging multiple measurements, the noise in the sensor signal can be reduced, leading to a more accurate measurement. If the right amount of white noise is added, the accuracy could be raised to 0.02%.

Overall, our refractometer provides a cost-effective and accurate solution for measuring sugar concentrations in various applications. Further research can be done to improve the resolution and accuracy even more.

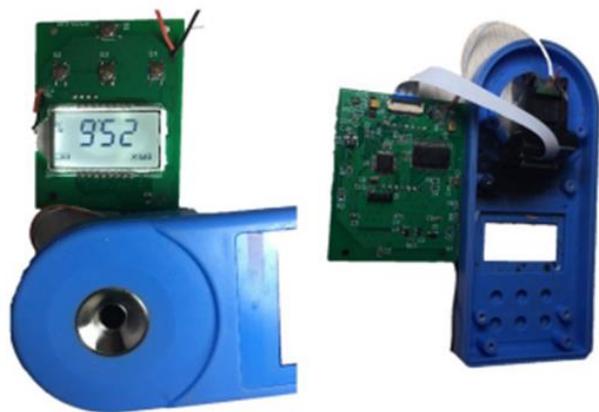

**Figure 12.** Prototype.